\documentclass[12pt]{article}
\usepackage[english]{babel}
\usepackage[latin1]{inputenc}
\usepackage[T1]{fontenc}
\usepackage{epsfig}
\usepackage{plain}

\newcommand{\un}{~\mathrm}
\newcommand{\ie}{{\em i.e. }}
\newcommand{\eg}{{\em e.g. }}
\newcommand{\unm}{~\mu\mathrm{m}}
\newcommand{\indice}[1]{\textnormal{\scriptsize{#1}}}
\newcommand{\indicito}[1]{\textnormal{\tiny{#1}}}

\begin{document}

\title{Origin of undesirable cracks during layer transfer}
\author{\hspace{0pt}L. Ponson$^{a)}$ , K. Diest, H.A. Atwater, \\
G. Ravichandran and K. Bhattacharya \\
\vspace*{-10pt} \\
\small{Division of Engineering and Applied Science, California Institute of Technology,} \\
\vspace*{-22pt} \\
\small{Pasadena, CA 91125, USA} \\
\vspace*{-2pt} \\
\small{$^{a)}$ Electronic mail: ponson@caltech.edu}}

\maketitle

\begin{abstract}
We investigate the origin of undesirable transverse cracks often observed in thin films obtained by the layer transfer technique. During this process, two crystals bonded to each other containing a weak plan produced by ion implantation are heated to let a thin layer of one of the material on the other. The level of stress imposed on the film during the heating phase due to the mismatch of thermal expansion coefficients of the substrate and the film is shown to be the relevant parameter of the problem. In particular, it is shown that if the film is submitted to a tensile stress, the microcracks produced by ion implantation are not stable and deviate from their straight trajectory making the layer transfer process impossible. However, if the compressive stress exceeds a threshold value, after layer transfer, the film can buckle and delaminate, leading to transverse cracks induced by bending. As a result, we show that the imposed stress $\sigma_\indice{m}$ \--- or equivalently the heating temperature \--- must be within the range $-\sigma_\indice{c} < \sigma_\indice{m} < 0$ to produce an intact thin film where $\sigma_\indice{c}$ depends on the interfacial fracture energy and the size of defects at the interface between film and substrate.
\end{abstract}

\section{Introduction}\label{Intro}
\quad Various applications in electronics and optics require the synthesis of high quality, defect-free single crystals on a substrate of a different material. Diverse heteroepitaxial growth processes have been proposed (\eg \cite{Ayers}), but these methods impose severe restrictions on the film/substrate combinations. Recently, the {\it layer transfer process} has been proposed and shows promise as an alternative when the film/substrate pair is very different \cite{Bruel, Diest}. The layer transfer is accomplished by implanting hydrogen or helium ions into a bulk crystal of the film to be synthesized and then bonding it to a substrate. Acting as damage precursors, these ions induce nucleation and growth of cavities when the specimen is heated at a sufficiently high temperature, transferring onto the substrate a single crystal thin film whose thickness corresponds to the depth of ion implantation. However, for some systems and some given heating conditions, undesirable transverse cracks are also produced in the thin film during the splitting process. This phenomenon renders the transferred thin film useless for applications in microelectronics and others. Therefore, understanding the origin of such cracks is crucial to avoid their formation. Identifying quantitatively the conditions and the systems that are advantageous to nucleate these undesirable cracks will help to define the limitations of the layer transfer process, and to design possible solutions to overcome these limitations. This motivates the present analysis and the mechanism of formation of these undesirable cracks is the central point of this study.

In Section \ref{Geometry}, the geometry used during the layer transfer process as well as the state of stress in the film are described. Then, a first possible origin of thin film failure is investigated in Section \ref{Stability}: the stability of cracks nucleating from defects introduced by ion implantation in the material to be cut is analyzed and we show that these cracks propagate parallel to the film/substrate interface only for a compressive state of stress in the film. In Section \ref{Section_Delam}, we show that a compressive stress in thin film can also lead to cracking by buckling, delamination and then failure of the film. This analysis provides a range for the compressive stress and therefore limitations of the heating temperature for a given system with fixed film thickness that will lead to a continuous thin film. In the following section, these theoretical predictions are combined with experimental observations made on a lithium niobate film bonded to a silicon substrate. The two failure mechanisms proposed previously to explain the presence of transverse cracks are clearly identified by a post-bonding analysis of the specimen after layer transfer. The theoretical criterion for good layer transfer ($-\sigma_\indice{c} < \sigma_\indice{m} < 0$) is found to agree with experimental observations.

\section{Geometry of the system and stress state of the film}\label{Geometry} 
\begin{figure}
\includegraphics[width=0.85\columnwidth]{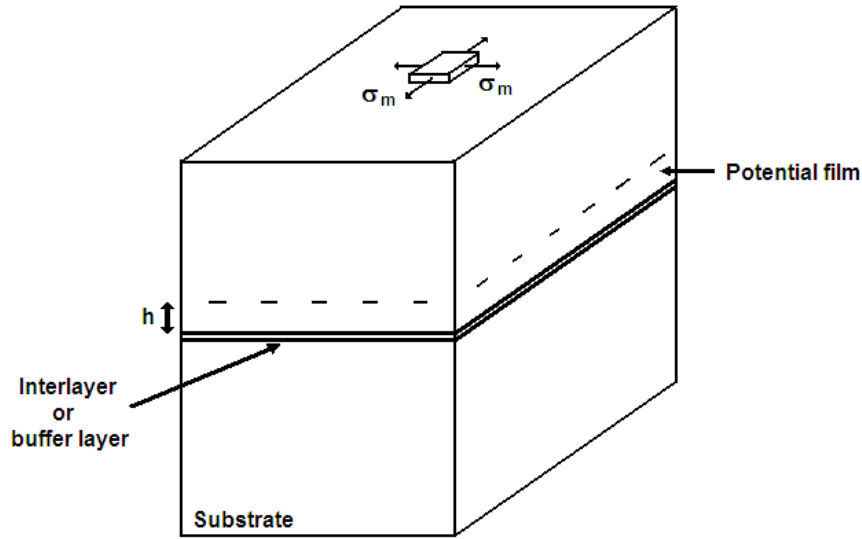}
\centering
\caption{Geometry and stress field of the layered system. The dashed plane coincides with the plane of ion implantation.}\label{Scheme}
\end{figure}

\quad To perform layer transfer, the material to be cut is bonded on a substrate as shown in Fig.\,\ref{Scheme}. A bonding layer, observed to improve adhesion and avoid undesirable cracking for some systems, is also shown. Its influence on the whole system is limited to the interface properties between film and substrate (fracture energy and defect size) so that this interlayer can be neglected in the following analysis without loss of generality. Such a layered system is then submitted to an elevated temperature $\Delta T$ and microcracks can nucleate in the plane of the film where hydrogen or helium has been previously implanted (dashed plane in Fig.\,\ref{Scheme}). When these microcracks coalesce, the bulk single crystal is separated from the transferred thin film with thickness $h$.

During the heating phase of the process, the film is submitted to an homogeneous bi-axial stress $\sigma_\indice{m}$ caused by the mismatch in thermal expansion between the film and the substrate. Noting $\Delta \alpha = \alpha_\indice{s}-\alpha_\indice{f}$, the difference between the linear thermal expansion coefficients of the substrate and the film, one can show that irrespective of the thickness and thermal properties of the bonding layer, the stress in the film is given by \cite{Freund2}
\begin{equation}
\sigma_\indice{m} = \frac{E}{1-\nu} \Delta T \Delta \alpha
\label{sigma_m}
\end{equation}
where $E$ and $\nu$ are the Young's modulus and the Poisson's ratio of the film, respectively. In the following sections, we will see that to ensure transfer of a thin film without undesirable transverse cracks, the stress imposed on the film must be within a certain range of values to be determined in the following sections.

\section{Stability of microcracks in the film}\label{Stability}
\begin{figure}
\includegraphics[width=0.6\columnwidth]{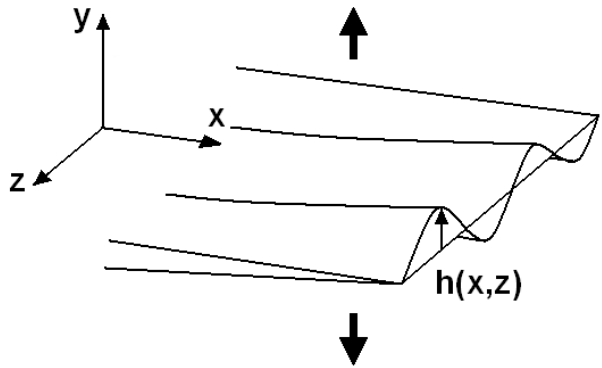}
\centering
\caption{Geometry of a slightly perturbed crack propagating in the film observed at sufficiently small scale 
so that the crack front appears on average to be straight.}\label{Out-plane}
\end{figure}

\quad Let's focus first on the trajectory of microcracks that initiate from the defects induced by the presence of hydrogen and helium in the specimen. To result in layer transfer, these microcracks are expected to propagate in a relatively straight manner, \ie parallel to the interface between the film and the substrate. The stability analysis of a 1D crack propagating in a 2D elastic medium submitted to an internal stress has been performed by Cotterell and Rice \cite{Cotterell}. To apply this result to the layer transfer process, we should make the hypothesis that the behavior of the 3D system as represented in Fig.\,\ref{Scheme} is analogous to that of a cut of the full system along a plane perpendicular to the film/substrate interface, \eg the plane (OYZ). In other words, we should suppose that the 2D penny-shaped microcracks propagating in the plane of ion implantation of the film can be approximated by 1D crack lines. This simplification is not obvious and in the following, we will study the more realistic situation of a 2D crack in a 3D elastic medium.
Figure \ref{Out-plane} represents a part of the crack front of a 2D penny-shaped microcrack when observed at a sufficiently small scale so that the crack front appears roughly straight, and parallel to the $z$-axis of the local coordinates (Oxyz) defined from the crack front geometry. While propagating along the $x$-direction, deflections of the crack front are generated by the heterogeneities of the film that can result from the damage and defects induced by ion implantation. Out-of-plane (along the $y$-axis) perturbations $h(x,z)$ as well as in-plane perturbations (along the $x$-axis) are generated. But one can show that for small deflections of the crack, only the out-of-plane perturbations are relevant to determine the local shearing at the crack tip and hence, the trajectory of the crack \cite{Larralde}. Therefore, only perturbations of the crack front along the y-axis have been represented in Fig.\,\ref{Out-plane}. To assess the stability of such perturbed cracks, one can determine if the deflection $h(x,z)$ will tend to zero or will diverge when the crack is propagating. To perform this analysis, we will apply the principle of local symmetry \cite{Goldstein, Cotterell, Hogdon} to the perturbed crack: locally, at every point of the front $M(x,h(x,z),z)$, the crack propagates in a pure mode I (opening) state of stress.
This condition can be written in the following way
\begin{equation}
K_\indice{II}(M(x,h(x,z),z) = 0.
\label{Eq_path}
\end{equation} 
Movchan {\it et al.} \cite{Movchan2} have calculated the mode II stress intensity factor of a slightly perturbed crack propagating in an infinite 3D elastic medium. Using their result, the local mode II stress intensity factor of cracks propagating in the ion implanted plane of the specimen can be expressed as
\begin{equation}
K_\indice{II} = \frac{K_\indice{I}^0}{2} \frac{\partial h}{\partial x}|_\indice{(x,z)}-\frac{K_\indice{I}^0}{2 \pi} \frac{2-3 \nu}{2- \nu} \int^{+ \infty}_{- \infty} \frac{h(x,z')-h(x,z)}{(z'-z)^2}dz' + \Delta K_\indice{II}^\indice{memory}
\label{KII_developpe}
\end{equation}
\noindent where the {\it memory} term $K_\indice{II}^\indice{memory}$ is given by
\begin{equation}
\begin{array}{lcl}
\vspace{5pt}
\Delta K^\indice{memory}_\indice{II}(x,z) & = & - \int_{-\infty}^x\int_{-\infty}^{+\infty} \Big\lbrace w^\indice{II}_\indice{x}(x-x',z-z')\left( \frac{\partial (hT_\indice{xx})}{\partial \indice{x}}|_\indice{(x',z')}+\frac{\partial (hT_\indice{xz})}{\partial z}|_\indice{(x',z')}\right)\\
\quad & + & w^\indice{II}_\indice{z}(x-x',z-z')\left( \frac{\partial (hT_\indice{xz})}{\partial x}|_\indice{(x',z')}+\frac{\partial (hT_\indice{zz})}{\partial z}|_\indice{(x',z')}\right) \Big\rbrace dx' dz'
\end{array}
\label{memory_term}\end{equation}
\noindent with 
\begin{equation}
\begin{array}{lcl}
\vspace{5pt}
w^\indice{II}_\indice{x}(x,z)& = & \frac{\sqrt{-2x} H(x)}{\pi^{3/2}*(x^2+z^2)}\left( 1+\frac{2\nu}{2-\nu}\frac{1-(z/x)^2}{1+(z/x)^2}\right)\\
w^\indice{II}_\indice{z}(x,z)& = & \frac{\sqrt{-2x} H(x)}{\pi^{3/2}*(x^2+z^2)}\frac{2\nu}{2-\nu}\frac{2z/x}{1+(z/x)^2}\\
\end{array}\label{Functions_w}
\end{equation}
where $H(x)$ is the Heaviside function. In the preceding expressions, $K_\indice{I}^0$ represents the average mode I stress intensity factor applied to the crack by the heated gas in the microcavities while $T_\indice{xx}$, $T_\indice{zz}$ and $T_\indice{xz}$ are the $T$-stress terms, or constant stresses imposed on the film in the absence of any crack. This implies that  $T_{xx} = \sigma_\indice{m}$, $T_\indice{zz} = \sigma_\indice{m}$ and $T_\indice{xz} = 0$. Equation (\ref{KII_developpe}) provides the different contributions to the mode II shearing at a point $M$ of the crack front induced by the perturbations of the fracture surface. The first term in Eq.\,(\ref{KII_developpe}) corresponds to the contribution of the local slope along the propagation direction, while the second term provides the shearing induced by perturbations of the crack front. The third term, also referred to as the {\it memory} term gives, as indicated by its name, the mode II contribution induced by the out-of-plane deviations of the crack line between its point of initiation and current position. This term is expressed as a function of the internal stress $\sigma_\indice{m}$ in the film, using the full expression of Eq.\,(\ref{memory_term}) and changing $T_\indice{xx}$, $T_\indice{zz}$ and $T_\indice{xz}$ by their relevant expressions.  Isolating the first term proportional to the local slope of the crack surface, the expression of the crack path, as given by the principle of local symmetry of Eq.\,(\ref{Eq_path}), can be rewritten as
\begin{equation}
\begin{array}{lcl}
\frac{\partial h }{\partial x}|_\indice{(x,z)} & = & \frac{1}{\pi} \frac{2-3 \nu}{2- \nu} \int^{+ \infty}_{- \infty} \frac{h(x,z')-h(x,z)}{(z'-z)^2}dz' \\
\quad & + & \sigma_\indice{m} \frac{2}{K_\indice{I}^0} \int_{-\infty}^x\int_{-\infty}^{+\infty} \left( w^\indice{II}_\indice{x}(x-x',z-z') \frac{\partial h}{\partial x}|_\indice{(x',z')} + w^\indice{II}_\indice{z}(x-x',z-z') \frac{\partial h}{\partial z}|_\indice{(x',z')} \right) dx' dz'.
\end{array}\label{Eq_stable}
\end{equation}

This equation predicts the path of crack evolution and can predict the stability of the failure process: if $\frac{\partial h}{\partial x} < 0 $, the local perturbation $h(x,z)$ is rapidly suppressed during crack propagation and the crack surface is on average flat. If $\frac{\partial h}{\partial x} > 0 $, even a small perturbation will grow and will lead to a macroscopic deviation of the crack plane from the horizontal plane of ion implantation (Oxz). In the latter case, crack propagation trajectory is referred to as unstable. This situation will clearly lead to catastrophic transverse cracks in the thin film during the layer transfer process.

Next, we assess the relevance of each term of the right-hand side of Eq.\,(\ref{Eq_stable}) that determines the stability of microcracks in the film during the heating process. The first term acts as a non-local restoring force along the crack front that tries to maintain it perfectly planar. However, it does not prevent the crack from deviating away from the mean crack plane \cite{Note1}, and therefore, does not contribute directly to the stability of the crack. The second term is composed of a part proportional to $\frac{\partial h}{\partial x}$ and another proportional to $\frac{\partial h}{\partial z}$.To assess the relative importance of each term, one can compare their two prefactors, $w^\indice{II}_\indice{x}$ and $w^\indice{II}_\indice{z}$, respectively. According to Eq.\,(\ref{Functions_w}), $w^\indice{II}_\indice{z}$ is smaller than  $w^\indice{II}_\indice{x}$ \cite{Note2}, and for most values of $(z,x)$, one gets $\frac{w^\indicito{II}_\indice{x}}{w^\indicito{II}_\indice{z}}\ll 1$. In other words, the stability of the crack is mainly dictated by the term proportional to $\frac{\partial h}{\partial x}$, leading to the approximation
\begin{equation}
\frac{\partial h }{\partial x}|_\indice{(x,z)} \simeq \sigma_\indice{m} \frac{2}{K_\indice{I}^0} \int_{-\infty}^x\int_{-\infty}^{+\infty} w^\indice{II}_\indice{x}(x-x',z-z') \frac{\partial h}{\partial x}|_\indice{(x',z')} dx' dz'.
\label{Eq_stable2}
\end{equation}
From this equation, one can assess the evolution of the local slope of the crack surface. From Eq.\,(\ref{Functions_w}), one notes that $w^\indice{II}_\indice{x}>0$. Therefore, the sign of $\sigma_\indice{m}$ will determine the evolution of the solution of Eq.\,(\ref{Eq_stable2}). If $\sigma_\indice{m}>0$, then $|\frac{\partial h }{\partial x}|$ is expected to increase when the crack propagates, while with $\sigma_m<0$, $\frac{\partial h }{\partial x}$ will tend to zero after a characteristic distance \cite{Note3}.

From the analysis of the stability of a crack propagating during the heating phase of the layer transfer process, one gets finally:
\begin{itemize}
\item[(i)] If the thin film is in a state of tensile stress ($\sigma_\indice{m}>0$), then the microcracks nucleated from the damage induced by ion implantation during the heating phase will deviate from their straight trajectory. One can therefore expect some difficulties obtaining good splitting of the upper part of the sample and some transverse cracks within the transferred thin film from systematic deviations of these microcracks.
\item[(ii)] If the film is in a state of compressive stress ($\sigma_\indice{m}<0$), then the microcracks are expected to propagate along a straight trajectory in a plane parallel to that of the ion implantation and will result in the transfer of an crack-free, single crystal thin film. This compressive stress state is obtained if the thermal expansion coefficient of the film if larger than that of the substrate (see Eq.\,(\ref{sigma_m})).
\end{itemize}

As a result, the condition $\sigma_\indice{m}<0$ is necessary to obtain straight crack propagation and therefore an intact thin film. Let us note that this result is not limited to multilayer systems and can be extended to other systems where the crack trajectory needs to be analyzed: a 2D crack will remain confined to a plane perpendicular to the external tensile loading if the stress is in tension along all the directions of this plane while it will deviate from the straight trajectory if the stress is compressive along the mean plane of the crack. This result extends the analysis of Cotterell and Rice \cite{Cotterell} limited to 2D systems to the more realistic situation of 3D systems. In the following section, we will investigate another possible origin of film cracking and show that there is a limit to the amount of compressive stress the film can support, and an excessively high compressive stress in the film can also lead to poor quality transferred thin films.

\section{Buckling, delamination and failure of the film}\label{Section_Delam}

\quad Here, another possible mechanism for film cracking during layer transfer process is investigated. Previously, we have shown that a state of tensile stress in the crystal containing the implanted plane must be avoided to ensure proper layer transfer. Therefore, systems with negative mismatch $\Delta \alpha = \alpha_\indice{s}-\alpha_\indice{f}$ between thermal expansion coefficients of the substrate and the film will be advantageously chosen. As an indirect consequence, the thin film freshly obtained after crystal layer transfer process might be submitted to a high compressive stress $\sigma_m<0$, as given by Eq.\,(\ref{sigma_m}).

It is well known that thin films under compression can buckle and delaminate \cite{Hutchinson, Freund2}. We will see that these processes can have catastrophic consequences because it can lead to film failure by bending. In the following section, the conditions leading to buckling, delamination and failure of the film produced by layer transfer and subjected to a compressive stress $\sigma_\indice{m}$ are investigated in detail. The film is supposed to be perfectly brittle so that the equations of elasticity for thin plates can be used. In addition, in first approximation, the fracture energy $G_\indice{c}$  of the film/substrate interface is assumed to be constant and independent of the phase angle $\phi = \textnormal{arctan}(\frac{K_\indicito{II}}{K_\indicito{I}})$ of the stress acting on the interface \cite{Freund2}.

\subsection{Delamination of a film with a semi-infinite defect}\label{Infinite}
\quad Under compressive stress, a film bonded to a substrate can delaminate in order to release its internal stress. For an infinite film bonded to an infinite substrate with a straight delamination front separating the film into two semi-infinite bonded and debonded parts, the elastic energy released during the propagation over a unit area is given by \cite{Freund2}
\begin{equation}
G_\indice{del} = |\sigma_\indice{m}^2| \frac{h}{2} \frac{1-\nu^2}{E}
\label{G_del}
\end{equation}
where $h$, $E$ and $\nu$ are the thickness, the Young's modulus and the Poisson's ratio of the film, respectively. Noting $G_\indice{c}$ the interfacial fracture energy between the film and the substrate \--- or between the film and the bonding layer if an additional layer has been added to the system [Fig.\,\ref{Geometry}], one can use the Griffith criteria $G_\indice{del} = G_\indice{c}$ providing the onset of crack propagation to get an expression of the critical stress $\sigma_\indice{del}$ for delamination
\begin{equation}
\sigma_\indice{del} = \sqrt{\frac{2 E G_\indice{c}}{h(1-\nu^2)}}.
\label{sigma_del}
\end{equation}
It must be emphasized that the initial condition taken here with a semi infinite debonded zone favoured interfacial crack propagation. In more realistic systems with defects or debonded zones of finite size at the interface between film and substrate (or bonding layer), such a level of compressive stress might not induce delamination. In addition, another mechanism must be taken into account to describe the delamination of films: buckling frequently observed in thin film under compression leads to modifications of the expression of the energy release rate $G$ as given in Eq.\,(\ref{G_del}). In the following section, we focus on this process and the conditions for film buckling. The out of plane displacements of the film are then taken into consideration in order to predict propagation of the delamination crack. In all the following, we limit our analysis to a 2D geometry of the specimen (\eg plane (Oyz) in Fig.\,\ref{Scheme}). We consider defects of length $2a$ at the interface between film and substrate, and determine if these debonded zones can grow and lead to catastrophic consequences for layer transfer, such as film failure.

\subsection{Buckling of the film}
\begin{figure}
\includegraphics[width=1.\columnwidth]{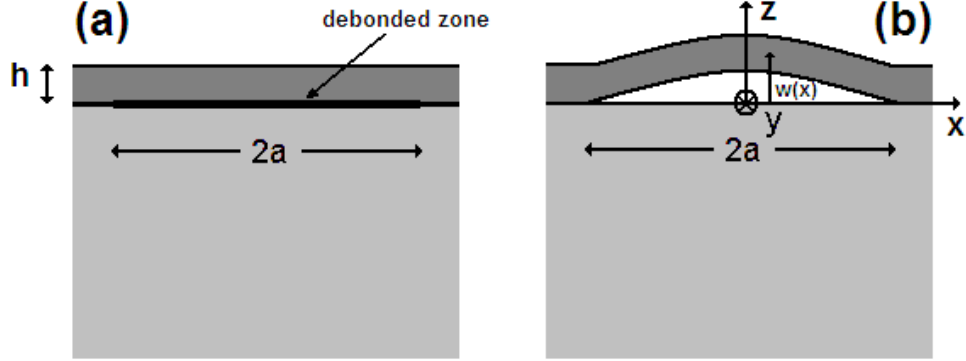}
\centering
\caption{Two-dimensional profile of a debonded part of a film without (a) and with buckling (b).}\label{Buckling}
\end{figure}

\quad We consider the situation represented in Fig.\,\ref{Buckling}(a) where an initial defect or debonded zone of size $2a$ is present at the interface between the film and the substrate. Submitted to a sufficiently high compressive stress, the film can buckle as represented in Fig.\,\ref{Buckling}(b), and a stability analysis of the film provides the expression of the critical stress \cite{Freund2}.

Considering now that the film is submitted to a given compressive stress $\sigma_\indice{m}$, one can use this expression to show that buckling will occur if the delamination zone is larger than a critical size $a_\indice{b}$ where
\begin{equation}
a_\indice{b} = \frac{\pi h }{2} \sqrt{\frac{E}{3 (1-\nu^2) |\sigma_\indice{m}|}}.
\label{a_b}
\end{equation}
This process is energetically favourable because in essence, it increases the effective length of the film.

\subsection{Propagation of the delamination front induced by film buckling}
\begin{figure}
\includegraphics[width=0.65\columnwidth]{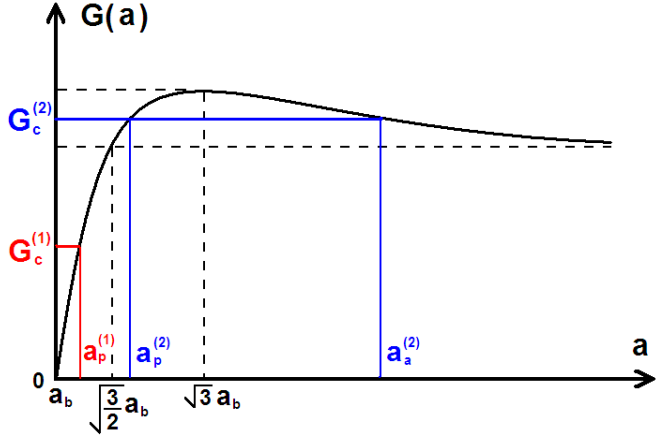}
\centering
\caption{Variations of the energy release rate of an interfacial crack at the edge of a buckled zone of length $a$ [Fig.\,\ref{Buckling}(b)]. $a_\indice{b}$ corresponds to the minimum length for a debonded zone in a film of the same thickness under the same compressive stress to buckle.}\label{G_a}
\end{figure}

\quad As mentioned previously, buckling of the film is affecting the energy release rate of the interfacial crack, so that the buckling pattern must be taken in consideration to predict the onset of delamination. In particular, the stress concentration at the edge of a debonded zone changes drastically with the size of the buckling zone. This effect is represented in Fig.\,\ref{G_a} where the variations of the energy release rate $G$ are represented as a function of the half-length $a$ of the debonded zone. For sufficiently large buckling zones, $G$ might reach $G_\indice{c}$ and the interfacial crack can propagate. To assess the critical size $a_\indice{p}$ that allows a buckling pattern to extend, one can derive the value of the energy release rate for a buckled zone of length $2a$ \cite{Freund2}
\begin{equation}
G(a) = \frac{\sigma_\indice{m}^2 (1-\nu^2) h}{2 E} \left( 1 - \frac{a_\indice{b}^2}{a^2}\right) \left( 1 + 3 \frac{a_\indice{b}^2}{a^2} \right)
\end{equation}
that is represented in Fig.\,\ref{Buckling}(b). It is interesting to note that at the onset of film buckling ($a=a_\indice{b}$), there is no driving force for delamination ($G=0$). However, if the compressive stress in the film is increased, the value of $a_\indice{b}(\sigma_\indice{m})$ will decrease leading finally to a net increase in the delamination driving force. As a result, propagation is possible at a certain stress level when the condition $G(a) = G_\indice{c}$ is satistied. One can express this criterion in terms of critical length $a_\indice{p}$ above which the buckling zone will extend
\begin{equation}
a_\indice{p} = \frac{\pi h}{2} \sqrt{\frac{E}{1-\nu^2}} \frac{1}{\sqrt{|\sigma_\indice{m}|}} \frac{1}{\sqrt{1+\sqrt{4-3\left(\frac{\sigma_\indice{del}}{\sigma_\indice{m}}\right)^2}}}
\label{a_p}
\end{equation}
\noindent where $\sigma_\indice{del}$ has been introduced in Eq.\,(\ref{sigma_del}). $a_\indice{p}$ being a decreasing function of $|\sigma_\indice{m}|$, it is also clear in this representation that a sufficiently large compressive stress will induce delamination. Note that Eq. (\ref{a_p}) is only valid for $|\sigma_\indice{m}|>\frac{\sqrt{3}}{2} \sigma_\indice{del}$. For smaller values of compressive stress $|\sigma_\indice{m}|$ in the film, the buckling zone remains stable regardless of the initial size of the debonded zone.

However, the previous analysis is limited to crack initiation and to predict the full evolution of the system beyond initiation, it is important to separate two cases, as illustrated in Fig.\,\ref{G_a}:
\begin{itemize}
\item[(1)] If the critical length $a_\indice{p}$ for interfacial crack propagation is smaller than $\sqrt{\frac{3}{2}} a_\indice{b}$, the equation $G_\indice{c} = G(a)$ has only one solution $a_\indice{p}$ given by Eq.\,(\ref{a_p}), corresponding to the size of the smallest defect leading to crack initiation. The condition $G_\indice{c} \geq  G(a)$ for crack propagation being always satisfied whatever $a>a_\indice{p}$, this situation corresponds to an unstable crack propagation without arrest of the crack.
\item[(2)] If the length $a_\indice{p}$ is larger than $\sqrt{\frac{3}{2}} a_\indice{b}$, the equilibrium equation for the debonding is satisfied for two crack lengths, $a_\indice{p}$ and $a_\indice{a}$. The elastic energy released is larger than the fracture energy only for crack extensions between these two lengthscales so that initiation and crack arrests occur successively for $a=a_\indice{p}$ (Eq. (\ref{a_p})) and $a=a_\indice{a}$ with
\begin{equation}
a_\indice{a} = \frac{\pi h}{2} \sqrt{\frac{E}{1-\nu^2}} \frac{1}{\sqrt{|\sigma_\indice{m}|}} \frac{1}{\sqrt{1-\sqrt{4-3\left(\frac{\sigma_\indice{del}}{\sigma_\indice{m}}\right)^2}}}
\label{a_a}
\end{equation}
\end{itemize}

The conditions for both situations can be rewritten in terms of stress, and unstable crack propagation corresponds to $|\sigma_\indice{m}| \geq \sigma_\indice{del}$ while crack arrest will be observed if $\sigma_\indice{del} > |\sigma_\indice{m}| \geq  \frac{\sqrt{3}}{2} \sigma_\indice{del}$. The value of the defect length corresponding to $a_\indice{p} = a_\indice{a}$ is noted $a_\indice{del}$ where
\begin{equation}
a_\indice{del} = \frac{\pi h}{2} \sqrt[4]{\frac{2 h E}{3 G_\indice{c} (1-\nu^2)}}.
\label{a_del}
\end{equation}

In both cases, the propagation of these interfacial cracks may adversely affect the quality of the transferred thin film. In particular, for sufficiently large buckled patterns, \ie large enough interfacial crack extension, a transverse crack induced by the bending generated in the film can fracture the crystal layer. It is worth to note that this process may not occur for an interfacial failure with a small extension. The conditions to obtain such transverse cracks are now discussed in detail. 

\subsection{Failure of the thin film induced by bending}
\quad When buckling occurs, the delaminated zone undergoes bending. If the original debonded zone is sufficiently small, bending increases while the size of the buckling zone increases. For a sufficiently large buckling zone, the film is not strong enough to support the tensile stress induced by bending in the film and a crack initiating from the upper surface of the film in $x=0$ will propagate parallel to the $y$-axis towards lower surface [Fig.\,\ref{Buckling}(b)]. In this geometry, crack propagation is expected to be highly unstable, and propagation will occur all through the crystal layer.

To predict the onset of crack initiation, we use a criterion based on the value of the curvature of the film (akin to critical strain), as \eg \cite{Audoly}: failure occurs when the curvature  $\frac{d^2w}{dx^2}$ at some point of the film exceeds the critical value $\frac{1}{R_\indice{c}}$ where $R_\indice{c}$ is a constant depending on the intrinsic strength of the material, but also on the state of surface of the freshly cut crystal. As clear from Fig.\,\ref{Buckling}(b), a possible transverse crack will initiate around $x=0$ where the local curvature of the film is maximum. The deflection $w(x)$ of the film is then expressed in terms of the delaminated zone size $2a$ and the compressive stress $\sigma_\indice{m}$  (\eg \cite{Freund2}), providing an expression for the maximum curvature $\frac{d^2 w}{dx^2}|_\indice{x=0}$ of the film. From this expression and the curvature based failure criterion introduced previously, one can show that transverse failure occur for buckled thin film of size larger than $a_\indice{f}$ with
\begin{equation}
a_\indice{f} = \pi \sqrt[4]{3(1-\nu^2)} \sqrt{h R_\indice{c}} \sqrt{\frac{|\sigma_\indice{m}|}{\sigma_\ell}} \sqrt{1-\sqrt{1- \left( \frac{\sigma_\ell}{\sigma_\indice{m}} \right)^2}}
\label{a_f}
\end{equation}
where
\begin{equation}
\sigma_\ell = \frac{E}{2 \sqrt{3(1-\nu^2)}} \frac{h}{R_\indice{c}}.
\label{sigma_l}
\end{equation}
Note that film failure is impossible if $|\sigma_\indice{m}|<\sigma_\ell$, regardless of the size of the debonded zone. For $|\sigma_\indice{m}|=\sigma_\ell$, we introduce the size $a_\ell=a_\indice{f}(\sigma_\ell)$ of the smallest debonded zone for which failure will occur
\begin{equation}
a_\ell = \pi \sqrt[4]{3 (1-\nu^2)} \sqrt{h R_\indice{c}}.
\label{a_l}
\end{equation}

\subsection{Comparisons of the various length scales of the problem and criterion for film failure}
\quad In the preceding paragraphs, the criteria for film buckling, extension of the debonded zone and transverse failure of the film were expressed in terms of debonded zone size. In these three cases, it was possible to define a critical size above which the process is expected to occur. For the specific case of propagation of the delamination front, our analysis showed also that above a critical length, the process will stop. These critical debonded sizes were shown to depend on the applied stress in the film, and their dependence with $\sigma_\indice{m}$ were explicitly given in Eq.\,(\ref{a_b}), (\ref{a_p}) and (\ref{a_f}).

\begin{figure}
\includegraphics[width=0.85\columnwidth]{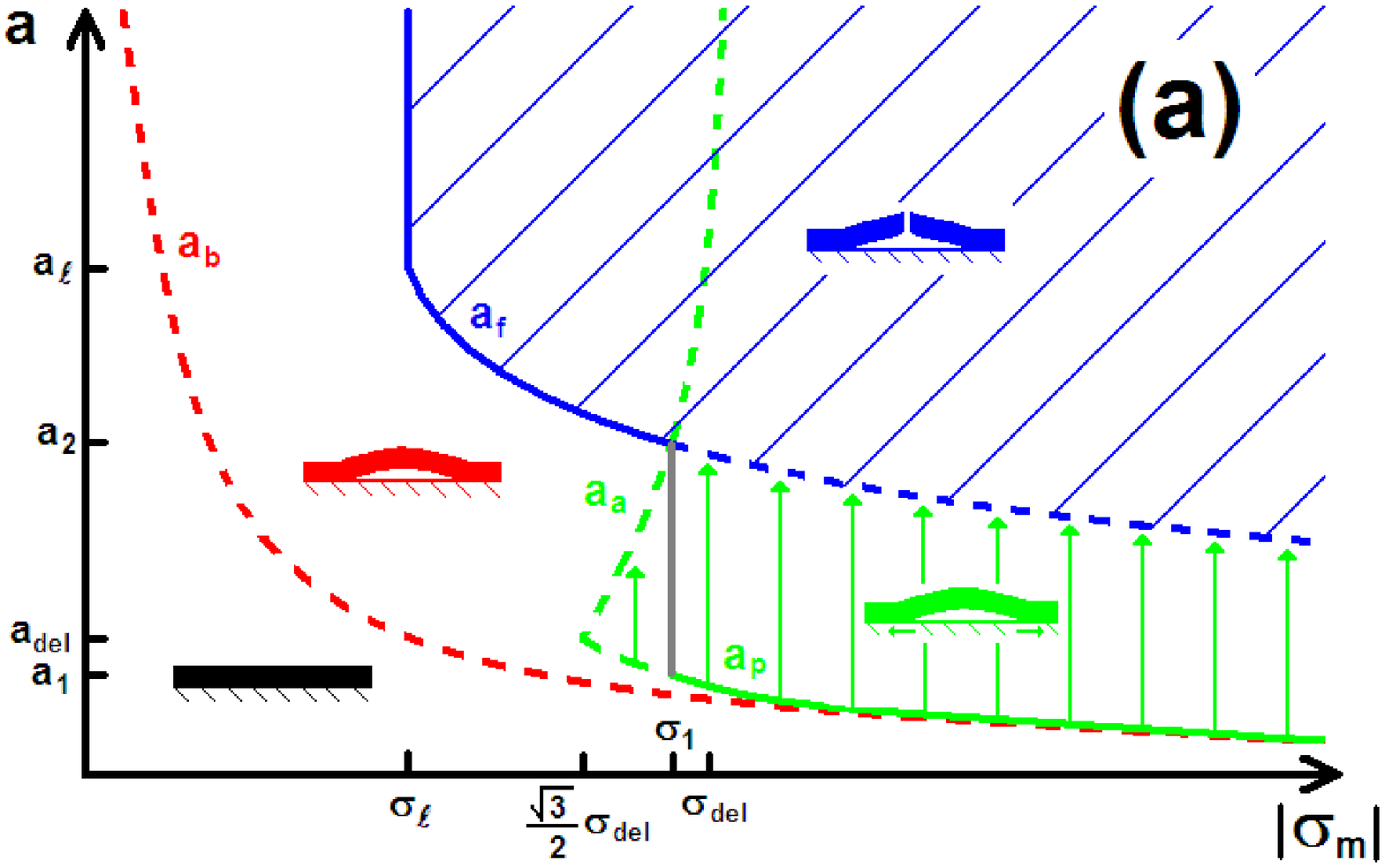}
\includegraphics[width=0.85\columnwidth]{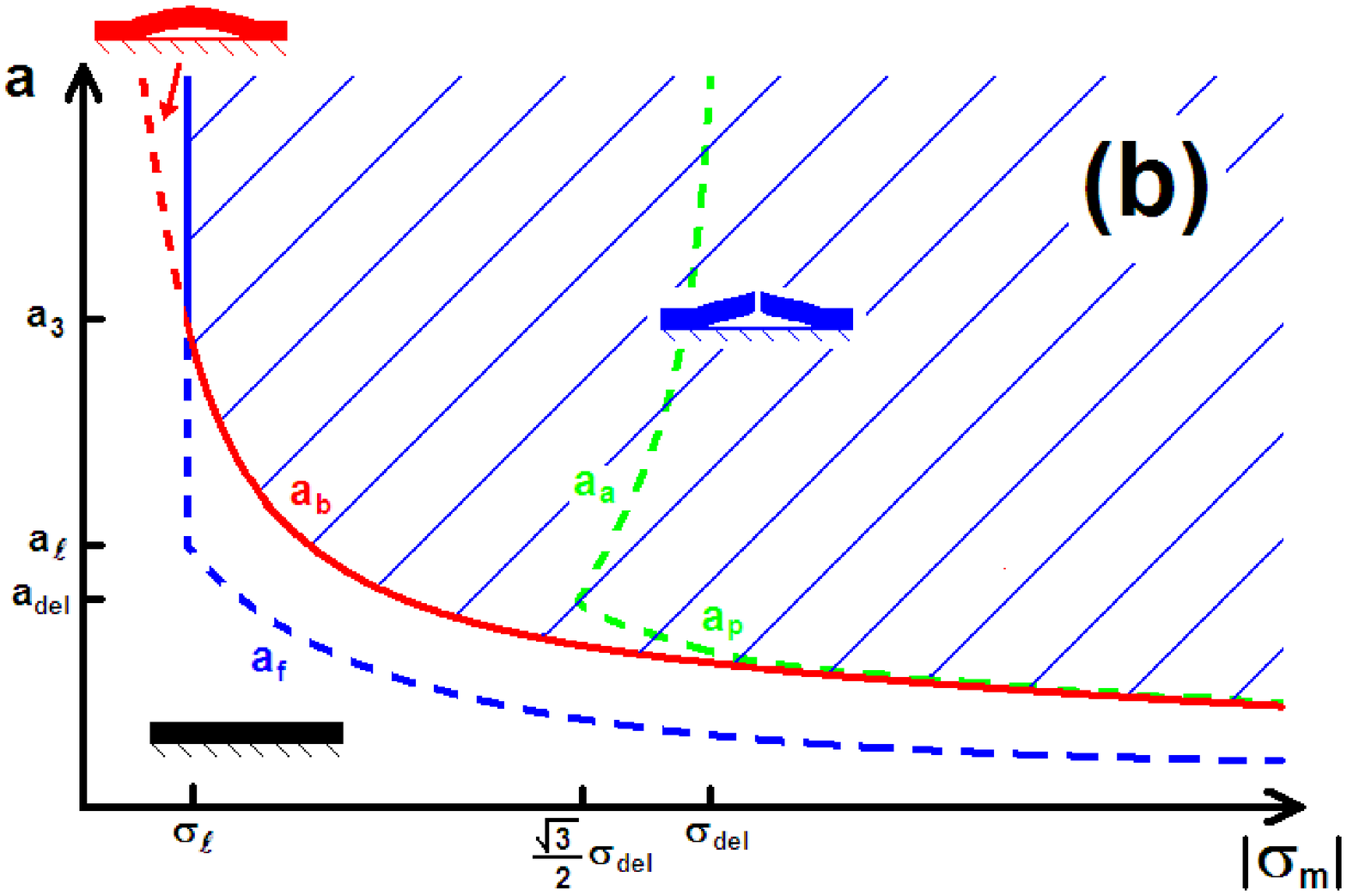}
\centering
\caption{Diagrams representing the state of the system and its evolution during the layer transfer process in two different cases: (a) film with high resistance to failure/small thickness; (b) film with low resistance/large thickness. In this representation, the state of the system is a point of coordinates ($|\sigma_\indice{m}|$, $a$) corresponding to the level of the compressive stress in the film and the half-length of the debonded zones at the interface film/substrate, respectively. Depending on the position on this graph, one can determine if the thin film will buckle (above the red line $a_\indice{b}(|\sigma_\indice{m}|)$), the delamination front will propagate (domain with vertical green arrows), or the film will break (hatched blue domain). To avoid film cracking during layer transfer, the system must remain in a state below the solid line in this representation.}\label{Phase_diagram}
\end{figure}

To be able to predict in a simple way film failure during layer transfer, these three criteria are represented on a same graph in Fig.\,\ref{Phase_diagram} where the compressive stress $|\sigma_\indice{m}|$ in the film is given along the abscissa and the half-length $a$ of the debonded zone along the ordinate. In this representation, the state of the system at a given time corresponds to a point of coordinates $(|\sigma_\indice{m}|, a)$. For each process studied, \ie buckling, delamination and film failure, the space $(|\sigma_\indice{m}|, a)$ can be divided into two distinct regions, separated by the curves $a_\indice{b}(|\sigma_\indice{m}|)$, $\{a_\indice{p}(|\sigma_\indice{m}|)$, $a_\indice{a}(|\sigma_\indice{m}|)\}$ and $a_\indice{f}(|\sigma_\indice{m}|)$. If the system, characterized by its coordinates $(|\sigma_\indice{m}|, a)$, is in the region defined for a given process, then this process will occur, while if the system corresponds to a point lower than the critical curve defined for the phenomenon, one does not expect this process to occur. Therefore, this relatively simple representation can be used to follow the temporal evolution of the layered specimen.

Such diagrams are represented in Fig.\,\ref{Phase_diagram} where the critical defect length $a_\indice{b}(|\sigma_\indice{m}|)$ for buckling, $a_\indice{p}(|\sigma_\indice{m}|)$, $a_\indice{a}(|\sigma_\indice{m}|)$ for propagation and arrest of the delamination front and $a_\indice{f}(|\sigma_\indice{m}|)$ for film failure are plotted. The relative position of the curve defining the domain for buckling and delamination is robust and independent to a greater extent of the specific value of the parameter of the problem. In particular, $a_\indice{b}$ is always smaller than $a_\indice{p}$, and in the limit of large compressive stress, $a_\indice{b} \simeq a_\indice{p}$. However, the position of the domain corresponding to film failure with respect to these curves may change with the value of the parameters. For illustrative purposes, two cases have been considered: on Fig.\,\ref{Phase_diagram}(a), $a_\indice{f}$ is larger than $a_\indice{b}$ and $a_\indice{p}$. This corresponds to large film thickness and/or highly resistant film. The other kind of systems corresponding to the diagram of Fig.\,\ref{Phase_diagram}(b) is associated with a small strength of the film \--- large critical film curvature $R_c$ at failure \--- and/or small film thickness.

In both diagrams, the hatched part corresponds to states of the system where the film is broken. Propagation of the interfacial crack, and thus extent of the debonded zone, is indicated by vertical arrows.

Let's consider at first the case of a highly resistant to failure film with a small thickness [Fig.\,\ref{Phase_diagram}(a)]. Whatever the initial size $a_\indice{ini}$ of the largest defects at the interface film/substrate, one can follow the evolution of the system during the layer transfer process. For example, let's take an initial defect size of the order of $a_\indice{del}$. During layer transfer, the temperature is increased and as a result, $|\sigma_m|$ also increases according to Eq.\,(\ref{sigma_m}). At the very beginning, the system evolution is represented by a horizontal line because the debonded zone remains unchanged. When the system reached the line $a_\indice{b}(|\sigma_\indice{m}|)$ demarcating the flat film and the film buckling, this zone starts to buckle, but $a$ still remains constant, so the specimen evolution can still be represented by an horizontal line. When the system reaches the line $a_\indice{p}(|\sigma_\indice{m}|)$ demarcating the stable buckled film and the propagation of the interfacial crack, there is delamination of the film and $a$ increases. Therefore, a vertical line now describes the evolution of the film geometry. Two cases are then possible: either the initial defect was sufficiently small ($a<a_\indice{1}$), and the extension of the buckled domain lead to film failure, the trajectory of the system in this representation reaching the border of the hatched zone. In that case, the critical debonded zone size before appearance of transverse cracks is provided by $a_\indice{p}(|\sigma_\indice{m}| \geq \sigma_\indice{1})$ (represented as a solid line). Or the crack stop before film failure, leading to a debonded zone of size $a_\indice{a}$ smaller than $a_\indice{f}$. The system will break only if the temperature is increased again, resulting to a quasi-static propagation of the delamination crack of half-length $a_\indice{a}$. Transverse cracks will eventually initiate if the compressive stress is sufficiently high so that $a_\indice{a}(|\sigma_\indice{m}|)$ reaches the critical size for failure $a_\indice{f}$. In this case, the critical compressive stress $\sigma_\indice{1}$ for film cracking is given by $a_\indice{a}(\sigma_\indice{1})=a_\indice{f}(\sigma_\indice{1})=a_\indice{2}$. Defining also $a_\indice{1}=a_\indice{p}(\sigma_1)$, one gets finally the following variations of the maximum admissible compressive stress $\sigma_\indice{c}$ with the initial defect size:
\begin{itemize}
\item[-] for $a_\indice{ini}<a_1$, $\sigma_\indice{c} = a_\indice{p}^{-1}(a_\indice{ini})$ where $a_\indice{p}(\sigma)$ is provided by Eq.\,(\ref{a_p}),
\item[-] for $a_1<a_\indice{ini}<a_2$, $\sigma_\indice{c} = \sigma_1$,
\item[-] for $a_2<a_\indice{ini}<a_\ell$, $\sigma_\indice{c} = a_\indice{f}^{-1}(a_\indice{ini})$ where $a_\indice{f}(\sigma)$ is provided by Eq.\,(\ref{a_f}),
\item[-] for $a_\ell<a_\indice{ini}$, $\sigma_\indice{c} = \sigma_\ell$ given in Eq.\,(\ref{a_l}).
\end{itemize}

Let's now focus on the case of films with large thickness and/or low resistance to failure. From the analysis of the corresponding diagram presented in Fig.\,\ref{Phase_diagram}(b), two cases can be isolated: for initial defects smaller than $a_3=a_\indice{b}(\sigma_\ell)$, the film remains intact as far as the critical stress for buckling is not reached. At this threshold, the debonded zone starts to buckle and a transverse crack appears at the same time. This means that the critical stress $\sigma_\indice{c}$ for film failure is provided by the expression of the buckling stress for a debonded zone of size $a_\indice{ini}$ that can be derived from Eq.\,(\ref{a_b}). For larger initial defects $a_\indice{ini}>a_3$, the film first buckles and then breaks when the compressive stress reaches the critical stress for failure $\sigma_\ell$. This leads us to conclude that
\begin{itemize}
\item[-] for $a_\indice{ini}<a_3$, $\sigma_\indice{c} = \frac{\pi^2}{12} \frac{E}{1-\nu^2}\left(\frac{h}{a_\indice{ini}}\right)^2$,
\item[-] for $a_3<a_\indice{ini}$, $\sigma_\indice{c} = \sigma_\ell$ given in Eq.\,(\ref{a_l}).
\end{itemize}

It is interesting to note that in the limit of very small defects $a \ll a_\indice{del}$, both kinds of systems represented by two rather different diagrams lead to the same expression of the critical compressive stress for film cracking. Using the approximation $a_\indice{b} \simeq a_\indice{p}$ valid for large compressive stress, one gets in both cases $\sigma_\indice{c} \simeq \frac{\pi^2}{12} \frac{E}{1-\nu^2}\left(\frac{h}{a_\indice{ini}}\right)^2$. The same remark is also valid in the limit of large defects for which $\sigma_\indice{c} = \sigma_\ell = \frac{E}{2 \sqrt{3(1-\nu^2)}} \frac{h}{R_\indice{c}}$ on a general manner.

This analysis provides an upper limit $\sigma_\indice{c}$ to the compressive stress that can be imposed on the film. Conversely using Eq.\,(\ref{sigma_m}), the maximum layer transfer temperature to which the system can be exposed to avoid failure can be also predicted. With the result obtained in Section \ref{Stability} from the stability analysis of microcracks leading to film splitting, one gets a range of admissible stress $-\sigma_\indice{c}<\sigma_\indice{m}<0$ for the system during the whole process, each limit corresponding to distinct failure modes. The theoretical predictions are compared with experimental observations in the following section.

\section{Discussion and comparison with experimental results}\label{Exp}
\quad To determine to what extent the previous analysis applies to experimental situations, two kinds of experiments for which transverse cracks in the film were observed have been analysed. Each one corresponds to one failure mechanism analysed in the previous sections. The first experiment is devoted to the study of the stability of microcraks in the film, and analyze the effect of the tensile/compressive state of stress on their trajectory. The second experiment has been designed to study the effect of large compressive stresses on the film.

\subsection{Effect of the compressive/tensile state of the stress on the stability of cracks}
\begin{figure}
\includegraphics[width=0.45\columnwidth]{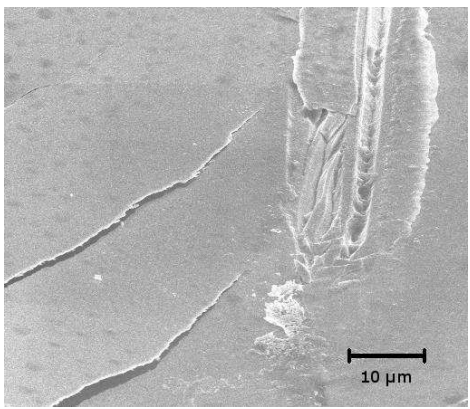}
\centering
\caption{SEM image of the top surface of an ion implanted $\mathrm{LiNbO}_3$ sample after heating. Transverse cracks can be seen coming from the implanted region in the $\mathrm{LiNbO}_3$, through the thin film, and emerging on the top surface of the sample.}\label{Crack_deviation}
\end{figure}

\quad For the first experiment, a sample of lithium niobate ($\mathrm{LiNbO}_3$), was implanted with hydrogen and helium to a depth of $h=400\un{nm}$ below the top surface. The specimen was simply heated and no bonding was involved. In this case, a coherent thin film of $\mathrm{LiNbO}_3$ is not separated from the rest of the material; rather, the cracks that initiated at the plane of implantation immediately deviate from a horizontal trajectory and finally emerge at the top surface of the sample [Fig.\,\ref{Crack_deviation}].

To explain these results, we assess the effect of the absence of substrate on the stress state in the $\mathrm{LiNbO}_3$ specimen: 
the stress $\sigma_\indice{m}$ remains equal to zero, even during the heating phase. Therefore, the cracks initiating from the implanted plane are unstable and as discussed in Section \ref{Stability}, they are expected to deviate from a horizontal trajectory. This observation is in agreement with the condition $\sigma_\indice{m}<0$ that was proposed in Section \ref{Stability} to ensure successful layer transfer.

\subsection{Effect of a high compressive stress on the film}
\begin{figure}
\includegraphics[width=0.45\columnwidth]{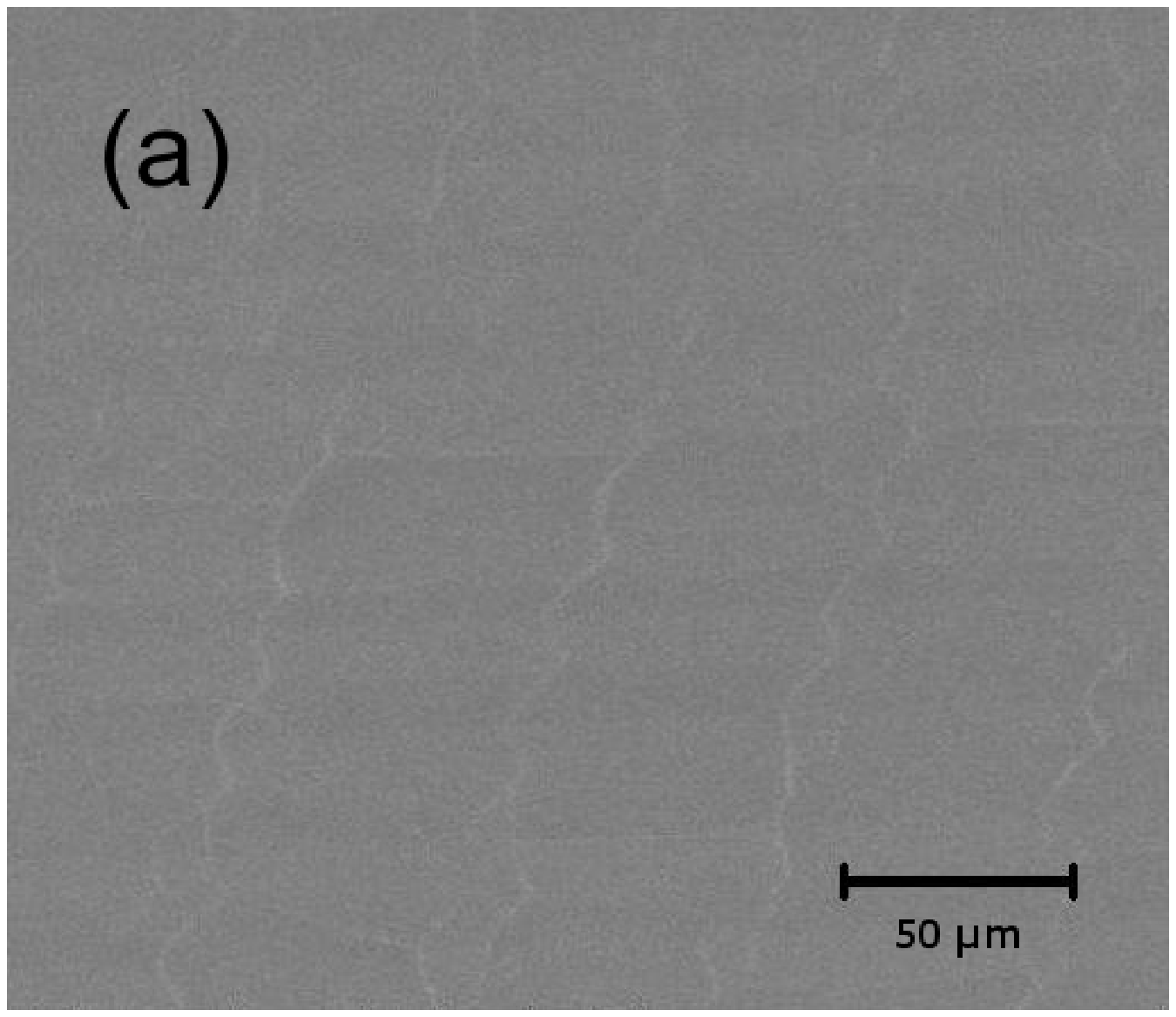}
\includegraphics[width=0.45\columnwidth]{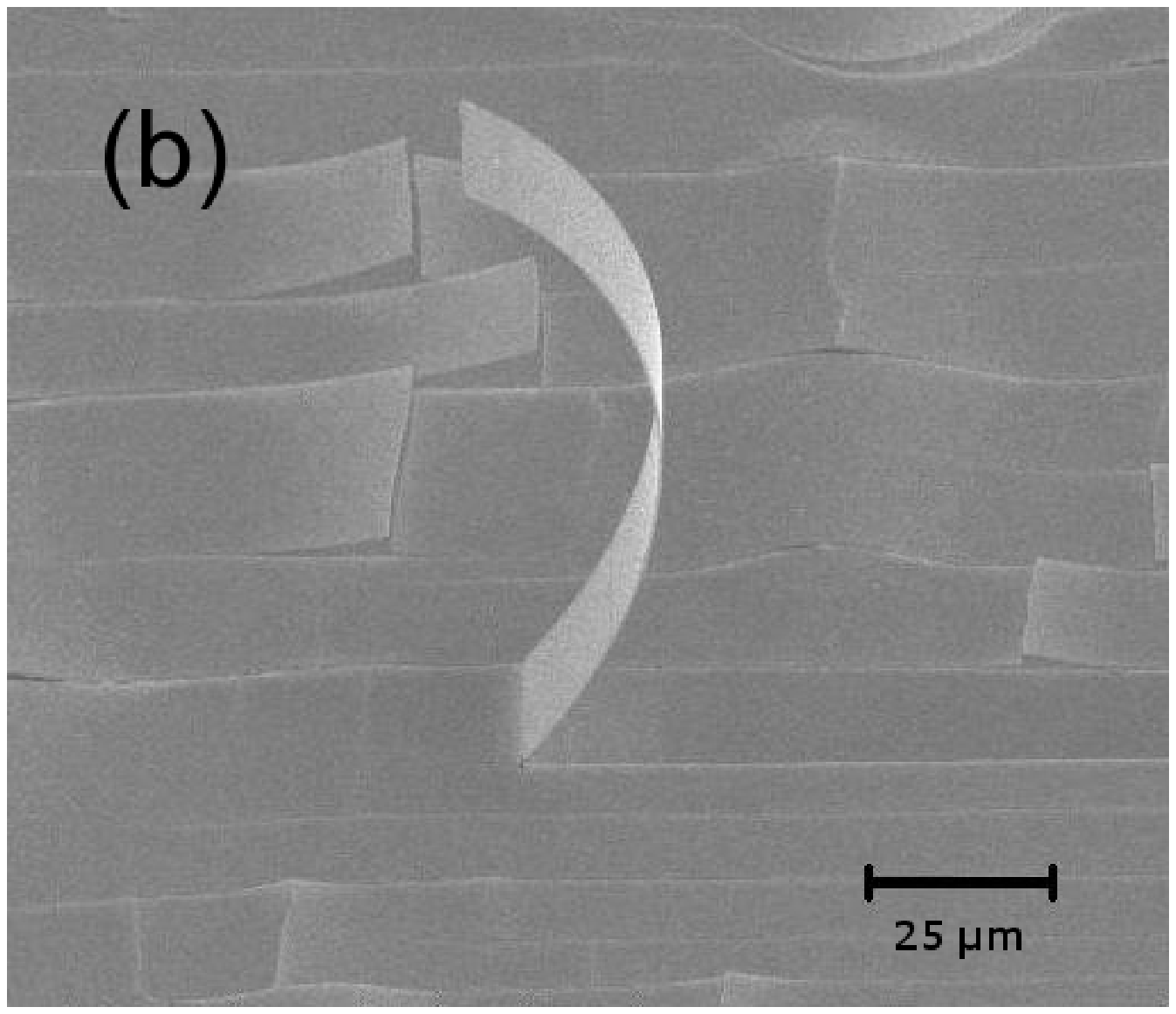}
\centering
\caption{SEM image of the free surface of the transferred thin film. (a) One can see a network of parallel fractures with "telephone cord" like cracks which are characteristic of buckling instabilities; (b) one can observe the network of secondary cracks perpendicular to the wavy cracks, also produced by buckling and failure of the film.}\label{Telephone_pattern}
\end{figure}

\quad The second experiment was performed on a system whose geometry corresponds to that represented in Fig.\,\ref{Scheme}. The crystal to be transferred is again ion implanted $\mathrm{LiNbO}_3$ with $h = 400\un{nm}$. The $\mathrm{LiNbO}_3$ (LNO) and silicon (Si) substrate were bonded together with minimal pressure and a silver bonding layer \cite{Diest2}. The substrate, the bonding layer, and the $\mathrm{LiNbO}_3$ specimen have square bases with sides of $1\un{cm}$. The thickness of the substrate and the bonding layer are $1 \un{mm}$ and $800\un{nm}$, respectively. The system is then heated roughly up to $750\un{K}$, so that $\Delta K \simeq 450\un{K}$, leading to a compressive stress in the film from the mismatch in thermal expansion coefficients, $\alpha_{LNO}$ being larger than $\alpha_{Si}$. In this case, layer transfer of the $\mathrm{LiNbO}_3$ specimen is obtained. This demonstrates that the compressive stress induced by the bonding of $\mathrm{LiNbO}_3$ onto a substrate with a smaller thermal expansion coefficient has enabled crack propagation along the plane of implantation.  This agrees with the predictions of Section \ref{Stability}. SEM images of the transferred $\mathrm{LiNbO}_3$ thin film indicate the presence of transverse cracks that have cut the film in various pieces [Fig.\,\ref{Telephone_pattern}]. One can see that these transverse cracks are all oriented in the same direction. This might correspond to the direction normal to the one of maximum thermal expansion coefficient of the $\mathrm{LiNbO}_3$ crystal \cite{Note4}. Also, these cracks are not straight, and follow a wavy trajectory, also referred as "telephone cord" like patterns, characteristic of thin film buckling \cite{Parry, Crosby}. This is strong evidence in support of the predictions of Section \ref{Section_Delam}: at first, the thin film buckles from a highly compressive stress, resulting in a network of buckling zones with a characteristic wavy geometry. Then, failure occurs by bending of the film where debonding has occured. This leads to transverse cracking in the film with the same wavy structure as the buckling.

An additional observation suggests that the transverse cracks observed in the film do not come from the deviation of microcracks which initiate at the plane of ion implantation, but result from buckling, delamination, and then failure of the film. The study of the other part of the sample ($\mathrm{LiNbO}_3$) that has been separated from the rest of the layered structure does not reveal any cracks on the freshly created surface. In other words, the interface between film and substrate plays a crucial role in the initiation of these undesirable cracks, while the ion implantation leads to a controlled splitting of the film, when bonded to a substrate with a smaller thermal expansion coefficient. This observation also suggests that a controlled splitting of the $\mathrm{LiNbO}_3$ single crystal is not enough to obtain a defect-free thin film, and the formation of transverse cracks by processes posterior to this splitting is also possible, as shown in Section \ref{Section_Delam}.

\begin{figure}
\includegraphics[width=0.65\columnwidth]{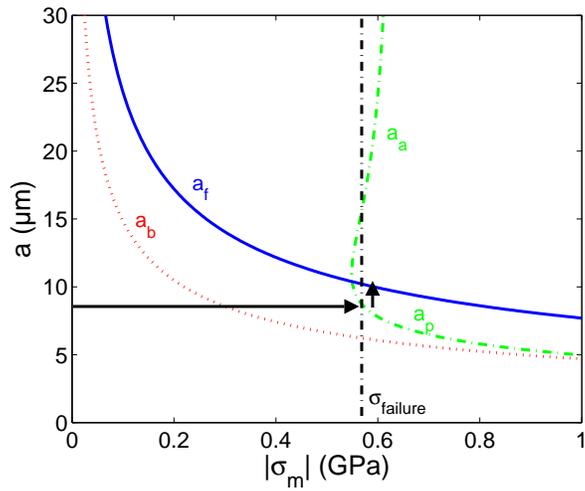}
\centering
\caption{Diagram representing the evolution of the $\mathrm{LiNbO}_3$/Ag/Si system during the layer transfer process. The vertical line is the experimental compressive stress $\sigma_\indice{failure} \simeq 0.57\un{GPa}$ in the film at failure, while the red dotted, green dashed and blue solid curves correspond to the critical values of the delaminated zone size for film bending, interfacial crack propagation/arrest, and film failure, respectively, as predicted by the theory. The evolution of the experimental system during layer transfer as expected from this diagramm is represented by the black arrows.}\label{Phase_diagram_exp}
\end{figure}

We now quantitatively compare the observations made in this experiment with the theoretical predictions made in Section \ref{Section_Delam}. In order to estimate the compressive stress at failure in the $\mathrm{LiNbO}_3$ film, Young's modulus and Poisson's ratio of $\mathrm{LiNbO}_3$ are taken to be $E = 150 \un{GPa}$ and $\nu = 0.32$, close to the values measured for similar materials \cite{Huntington}. The thermal expansion coefficients of Si and $\mathrm{LiNbO}_3$ are $\alpha_{Si} = 2.6 \times 10^{-6}\un{K^{-1}}$ and  $\alpha_{LNO} = 8.2 \times 10^{-6}\un{K^{-1}}$ \cite{Kim}, respectively, leading to $\Delta \alpha = - 5.6 \times 10^{-6}\un{K^{-1}}$. The critical radius of curvature for film failure under bending is estimated to be $R_\indice{c} \simeq 1\un{cm}$. Even though this value is a rather rough estimate, it is important to note that the shape of the curve $a_\indice{f}$ is rather insensitive to the value of $R_c$ in the range of interest $\sigma_\indice{m}>0.1\un{GPa}$ \cite{Note5}. Ceramic materials that are bonded to silver layers exhibit fracture energies on the order of $G_\indice{Ic} \simeq 1-2\un{J.m^{-2}}$. In the following, we have kept the fracture energy as a free parameter and chosen the value that enables the best agreement between experimental observations and theoretical predictions. The value so obtained is then compared with the expected values for ceramic-silver fracture energy.

Using the previous numerical values and Eq.\,(\ref{sigma_m}), it is possible to estimate the compressive stress $\sigma_\indice{failure} \simeq 0.57\un{GPa}$ in the film at $T \simeq 750\un{K}$ for which undesired cracks appear. For the $\mathrm{LiNbO}_3$/Ag/Si system studied here, one can also calculate the failure diagram to determine the state of the system with respect to $|\sigma_\indice{m}|$ and $a$ [Fig.\,\ref{Phase_diagram_exp}]. To reproduce corretly the experimental observations, one chooses $G_\indice{Ic} \simeq 0.5\un{J.m^{-2}}$ that is smaller but comparable to the expected values $G_\indice{Ic} \simeq 1-2\un{J.m^{-2}}$. The diagram so obtained is analogous to Fig.\,\ref{Phase_diagram}(a) plotted in a general case. The value of the compressive stress at $T \simeq 750\un{K}$ is also represented on this diagram as a vertical dashed line. It is now possible to identify the different processes that have led to the failure of the film. Using the representation of the system state shown in Fig.\,\ref{Phase_diagram_exp}, the initiation of the transverse cracks in the film is given by the intersection of the vertical dashed line giving the level of stress at film failure with the curve $a_\indice{f}(|\sigma_\indice{m}|)$ demarcating intact films from broken films. This provides a raisonable estimate $a \simeq 8\unm$ of the size of the defects at the interface between the silver bonding layer and the $\mathrm{LiNbO}_3$ film that will lead ultimately to undesirable cracks in the film.

From this diagram, one can also follow the history of the film failure during the heating phase. The evolution of the system during the initial phase is described by the horizontal arrow represented in Fig.\,\ref{Phase_diagram_exp}. One observes at first that the defects at the interface between Ag and $\mathrm{LiNbO}_3$ of size $a \simeq 8\unm$ will start to buckle for $\sigma_\indice{m} \simeq 0.3\un{GPa}$ (corresponding to a temperature of $\simeq 450\un{K}$). This value is given by the intersection of the horyzontal arrow with the curve $a_\indice{b}$. When the compressive stress in the film is sufficiently high, close to $\sigma_\indice{failure}$, the interfacial cracks start to propagate. A network of debonded zones with a telephone cord like geometry then develops. This process will lead ultimately to the telephone cord like cracks observed post-mortem on the thin film surface [Fig.\,\ref{Telephone_pattern}(a)] when the debonded zones will start to extend in the transverse direction \cite{Note6}. The evolution of the system in this last regime is described by the vertical arrow represented in Fig..\,\ref{Phase_diagram_exp}. Finally, the failure of the film corresponding to the intersection of the vertical arrow with the curve $a_\indice{f}$ is obtained for a lateral size of the debonded zones of the order of $a \simeq 10\unm$. This is fully compatible with the post-mortem observations made on the film surface after layer transfer.

\section{Conclusion}
The origin of the undesirable cracking often observed during layer transfer has been investigated. From our theoretical analysis based on Fracture Mechanics, it appears that the state of stress in the film, direct consequence of the mismatch between the thermal expansion coefficients of the film and the substrate, is driving the failure processes. More precisely, two phenomena identified in experimental examples are studied in detail and shown to induce catastrophic failure of thin films obtained by layer transfer: (i) the microcracks that are made to propagate in the implanted plane parallel to the film/substrate interface to split the specimen can deviate from their horizontal trajectory and cut the film. The analysis of their stability in the full 3D geometry of the considered system shows that these microcracks will not follow a straight path if the film is submitted to a compressive stress $\sigma_\indice{m}<0$; (ii) an important tensile stress in the film can also have catastrophic consequences. When the specimen is already cut but still heated, defects at the film-substrate interface can buckle and induce film delamination, resulting ultimately in a failure of the film by bending. This process has been analyzed in detail and the critical stress (critical temperature) at which each stage occurs has been expressed in term of defect size, film thickness and fracture properties of the film. Therefore, it is possible to predict the maximum compressive stress $\sigma_\indice{c}$ that can be sustained by the system. Taking into consideration both these failure processes, one can define a range of admissible stresses $ - \sigma_\indice{c} < \sigma_\indice{m} < 0$ in the film.
 
From these results, it is now possible to identify the systems amenable to the layer transfer technique. In particular, the conditions on the admissible stress in the film can be expressed in terms of system properties: the substrate must be chosen so that its thermal expansion coefficient is smaller than that of the film. But this condition is not sufficient and above a critical heating temperature corresponding to a compressive stress $\sigma_\indice{c}$, cracked film will be produced. This temperature must be smaller than the one necessary to make the microcracks propagate in the implanted plane of the film. To overcome this difficulty and increase the admissible stress in the film, the quality of the interface between film and substrate must be improved, decreasing both the defect size and increasing the interfacial fracture energy. A plastic interlayer (\eg Ag, Pt) used to accommodate the contact between the both surfaces might be relevant. Increasing the film thickness might be also an alternative to avoid failure of the system.

Finally, let us note that another effect may result in intrinsic limitations of the layer transfer process. The material to be cut is usually a single crystal with preferential cleavage planes. As a result, the implanted plane may not correspond to the easy direction of the film. To what extent a crack can propagate in the direction of maximal tensile stress \--- parallel to the substrate/film interface \--- rather than deviate for a plane of lower fracture energy is still a matter of debate \cite{Buczek, Azdhari}. A theoretical framework providing accurate predictions on the propagation direction of cracks in anisotropic media would lead to a clear determination of the systems as well as the crystallographic film orientations that can be obtained using the layer transfer technique. Works are currently in progress in this direction.

\vspace{20pt}

\noindent {\large {\bf acknowledgment}}

This work has been supported by the Center for the Science and Engineering of Materials (CSEM).

\end{document}